\documentclass[structabstract]{aa}
\usepackage{graphicx}
\usepackage{amsmath}
\usepackage{amssymb}

\def\be{\begin{equation}}
\def\ee{\end{equation}}
\def\bea{\begin{eqnarray}}
\def\eea{\end{eqnarray}}
\newcommand{\msun}{\rm{M}_\mathrm{\rm \sun}}

\begin{document}

\title{Calibration of radii and masses of open clusters with a simulation.}


\author{
A. Ernst\inst{1}
\and
A. Just\inst{1}
\and
P. Berczik\inst{1,2,3} 
\and 
M. I. Petrov\inst{4}
}

\institute{
Astronomisches Rechen-Institut, Zentrum f\"ur Astronomie der Universit\"at Heidelberg, 
M\"onchhofstrasse 12-14, D-69120 Heidelberg, Germany 
\and
National Astronomical Observatories of China, Chinese
Academy of Sciences, Datun Lu 20A, Chaoyang District, Beijing 100012, China 
\and 
Main Astronomical Observatory, National Academy of
Sciences of Ukraine, Akademika Zabolotnoho 27, 03680 Kyiv, Ukraine 
\and 
Institut f\"ur Astronomie der Universit\"at Wien, T\"urkenschanzstra{\ss}e 17, A-1180 Wien, Austria
}

\date{Received ... Accepted ...}

\abstract{A recent new approach to apply a simple dynamical 
mass estimate of tidally limited star clusters is based on the identification of the tidal 
radius in a King profile with the dynamical Jacobi radius. The application to an unbiased 
open cluster catalogue yields significantly higher cluster masses compared to the 
classical methods.}{We quantify the bias in the mass determination as function of 
projection direction and cluster age by analysing a simulated star cluster.}{We use 
direct $N$-body simulations of a star cluster including stellar evolution
in an analytic Milky Way potential and apply a best fit to the projected number 
density of cluster stars.}{We obtain significantly overestimated star cluster masses 
which depend strongly on the viewing direction. The overestimation is typically 
in the range of 10-50 percent and reaches a factor of 3.5 for young clusters. Mass
segregation reduces the derived limiting radii systematically.}{}



\maketitle

\section{Introduction}

In a series of papers (Piskunov et al. 2007, 2008a,b) a new approach to determine the 
masses of open star clusters (OCs) was developed and used to determine the initial 
and present day mass function of OCs in the solar neighbourhood.
The new method is based on the determination of the tidal radius $r_t$ from the cumulative 
number of cluster members as function of projected distance to the cluster center. 
For each cluster the tidal radius $r_t$ is determined from projected number density 
profiles by fitting a King 1962 profile (King 1962). 
The identification of the King cutoff radius  $r_t$ with the ``Jacobi'' radius  $r_J$ 
(i.e. the dynamical tidal radius, which is the distance from the 
cluster center to the Lagrange points $L_1$ and $L_2$) yields
then the OC mass from the standard formula (Equation \ref{eq:rjac} below solved 
for $M_{\rm cl}$). The 
application of this dynamical mass estimate of tidally limited clusters to an unbiased OC 
catalogue yields an independent mass determination compared to the classical methods. A detailed comparison with other methods of cluster mass determinations 
is also given. 
In a second step the method is extended to all OCs of an unbiased cluster catalogue 
by establishing a transformation of the observed semi-major axis and central surface 
density to $r_t$. 
 These results were then used to derive the cluster present day mass function (CPDMF) 
 and the initial mass function of OCs (CIMF) in the extended solar neighbourhood. 
 Adopting a constant cluster formation rate over the last 10\,Gyr yields a surface density of 
 $18 \, \msun\mbox{pc}^{-2}$ of stars born in OCs. 
 This corresponds to a fraction of 37\% of disc stars which were born in OCs (R\"oser et al. 2010). 
 This is large compared to the classical values of the order of 10\% or less 
 (e.g. Wielen  1971, Miller \& Scalo 1978).

Some crucial assumptions enter the dynamical mass determination based on fitting a 
King profile: a) The OC fills its Roche lobe 
in the tidal field of the Milky Way. For compact (e.g. Roche-lobe underfilling)  clusters 
$r_J$ and as a consequence the mass can be underestimated by a large amount. 
b) The effect of mass segregation can be neglected, i.e. star counts of the upper 
main sequence, which dominate the observed cluster members, are representative 
for the mass distribution. c) The elliptic shape of the clusters  and the contamination 
through tidal tail stars do not result in a systematic bias. Shape parameters 
were measured by Kharchenko et al. (2009) and the distribution of tidal tail stars 
were investigated in detail (e.g. Just et al. 2009).
d) The tidal radius $r_t$ determined by fitting the cumulative projected mass profile 
represents the Jacobi radius $r_J$ to derive the cluster mass. Since the cluster mass 
depends on the third power of $r_J$, the method is very sensitive to systematic errors 
in the derivation of $r_J$.

In the present paper we quantify the possible bias introduced by the identification of 
the tidal radius from a King profile fitting $r_t$ with the Jacobi radius  $r_J$ used for 
the mass determination by Piskunov et al. (2007). 
We apply the King profile fitting procedure to a direct $N$-body simulation of a dissolving 
star cluster at different evolutionary states.
We have simulated a star cluster on a circular orbit at $R_C=8.5$ kpc which
evolved in the tidal field of the Milky Way including stellar evolution. We took 
snapshots of the evolved
model with all stellar masses and positions at four different times and projected
the snapshots from the perspective of an observer on Earth 
(at $R_0=8$ kpc) onto the sky, at different positions along its orbit. 
After all, we determined the model's
limiting radius $r_t$ by fitting the projected cumulative mass profile
with Equation (\ref{eq:pcm}) and compared $r_t$ to the actual Jacobi radius $r_J$.

The paper is organized as follows: In Section 2 we discuss the method of $N$-body
simulations in the external potential of the Milky Way, Section 3 contains the theory of 
the cluster geometry in a tidal field. In Section 4, we show a simple iterative method
to determine the Jacobi radius of an $N$-body model of a star cluster in a tidal field. 
In Section 5 the projection and fitting methods are described. 
Finally, Section 6 contains the results and section 7 the conclusions.

\section{Numerical simulation}

We analyse in detail a numerical simulation of a star cluster with initial mass 
$M_0=10^4\msun$ on a circular orbit in an analytic Milky Way potential. It is the 
fiducial cluster simulation (run 10) discussed in Just et al. (2009). We have chosen 
this cluster, because it is a typical representative for the high-mass end of the 
observed OCs. Since the total mass of the cluster system is dominated by the 
high-mass end, the correction of biases in the mass determination are most 
important in that parameter regime. The cluster is set up as a $W_0=6$ King 
model with a half-mass radius of 8\,pc. The extension of the cluster initially 
exceeds the Roche lobe initially leading to an enhanced mass loss in the 
first 0.5\,Gyr. We used a Salpeter IMF and included mass loss by stellar 
evolution. The total lifetime of the cluster at a circular orbit with $R_C=8.5$\,kpc 
is 6.3\,Gyr. For details of the evolution see Just et al. (2009).

\begin{table}
\caption{The list of galaxy component parameters. The first column gives the
component, the second the mass, and the third and fourth the
Plummer-Kuzmin parameters (equation~\ref{eq:eq-gal}).}
\label{tab:gal-par}
\begin{center}
\begin{tabular}{lcrr}
\hline\noalign{\smallskip}
Component & M [M$_\odot$] & $a~[{\rm kpc}]$ & $b~[{\rm kpc}]$ \\
\noalign{\smallskip}
\hline
\noalign{\smallskip}
 Bulge & $1.4 \times 10^{10}$ & 0.0 &  0.3 \\
 Disk  & $9.0 \times 10^{10}$ & 3.3 &  0.3 \\
 Halo  & $7.0 \times 10^{11}$ & 0.0 & 25.0 \\ 
\hline
\end{tabular}
\end{center}
\end{table}

\begin{figure}
\includegraphics[angle=90,width=0.5\textwidth]{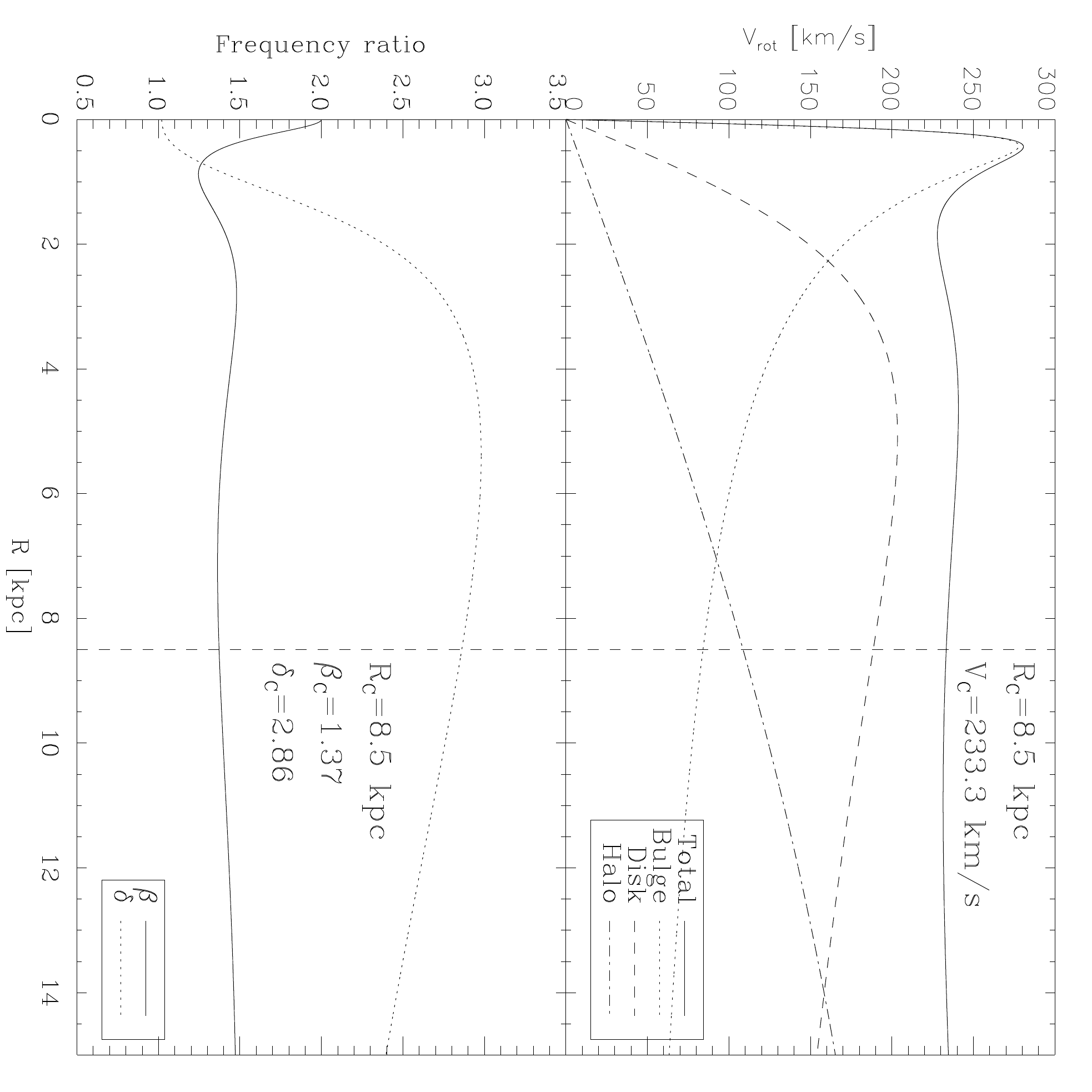} 
\caption{Top: Rotation curve (at $z=0$) of the 3-component Plummer-Kuzmin 
model of the Milky Way.
Bottom: Epicyclic and vertical frequency parameters $\beta=\kappa/\Omega$ and $\delta = \nu/\Omega$ (at $z=0$).} 
\label{fig:frequ}
\end{figure}

For the high-resolution simulation of a dissolving star cluster with $N=40404$ particles
in the tidal field of the Milky Way the direct $N$-body code $\phi${\sc grape}\footnote{The present 
version of the code is publicly available 
from one of the authors FTP site: 
{\tt ftp://ftp.ari.uni-heidelberg.de/staff/berczik/ 
phi-GRAPE-cluster/code-paper/}.} (Harfst et al. 2007) 
has been used in combination wit the micro-{\sc grape6} 
special-purpose hardware at the Astronomisches Rechen-Institut (ARI) 
in Heidelberg\footnote{GRACE: {\tt http://www.ari.uni-heidelberg.de/grace}}. $\phi${\sc grape} is an acronym for Parallel Hermite Integration
with {\sc grape}. The code is written in {\sc ansi-c} and uses a fourth-order Hermite 
scheme (Makino \& Aarseth 1992) for the orbit integration. 
It is parallelized and uses the MPI library for communication between the processors.
The force computations are executed on the fast special-purpose hardware {\sc grape}.
The special-purpose micro-{\sc grape6} hardware cards are especially designed to calculate 
gravitational forces in $N$-body simulations very fast using parallelization with 
pipelining (see Harfst et al. 2007 and references therein).

The code $\phi${\sc grape} does not use regularization as the codes {\sc nbody4} or
{\sc nbody6++} (Aarseth 1999, 2003;  Spurzem 1999) but a standard Plummer type 
$N$-body gravitational softening. The softening length in the 
model used for the current work was $\epsilon=10^{-3}$ pc. We tested with different 
softening lengths $\epsilon= 10^{-3}, 10^{-4}$ and $10^{-5}$ pc that there are no 
significant differences regarding shape evolution and star cluster mass loss.

For the simulation of a star cluster in the tidal field of the 
Galaxy the $N$-body problem is solved in an analytic background potential. 
We use an axi-symmetric 3-component model, where bulge, disc and halo are 
described by Plummer-Kuzmin models (Miyamoto \& Nagai 1975) with
the potential

\be
\Phi(R,z) = - \frac{ GM }{ \sqrt{R^2 + (a + \sqrt{b^2 + z^2} )^2} }. \label{eq:eq-gal}
\ee

\noindent
The parameters $a, b$ and $M$ of the Milky Way model are given in Table \ref{tab:gal-par}
for the three components.

The top panel of Figure \ref{fig:frequ} shows the rotation curve of the 3-component model of the Milky Way.
The parameters of the 3-component model
are chosen such that the rotation curve matches that of the Milky Way (Dauphole \& Colin 1995). 
At the solar radius $R_0=8.0$ kpc, which was assumed in this study, 
the value of the circular velocity is 
$V_0=234.2$ km/s. The values of Oort's constants $A$ and $B$ are consistent 
with the observed values $(A,B) = (14.5 \pm 0.8, -13.0 \pm 1.1)$ km/s/kpc derived by
Piskunov et al. (2006). More generally, the dimensionless epicyclic and vertical 
frequency parameters are given by

\bea
\beta^2 &=& \kappa^2/\Omega^2 = 2 \left( \frac{d\ln\Omega}{d\ln R} + 2 \right) \ \ \ \mathrm{and} \\
\delta^2 &=& \nu^2/\Omega^2 = \frac{4\pi G\rho}{\Omega^2} + 2 - \beta^2 \label{eq:delta2}
\eea

\noindent
where $\kappa$, $\nu$ and $\Omega$ are the epicyclic, vertical and circular frequencies of a near-circular orbits and $\rho$ is the local galactic density (see Oort 1965 for the derivation of $\delta^2$).
The bottom panel of Figure \ref{fig:frequ} shows the course of the epicyclic and vertical frequency parameters
$\beta$ and $\delta$. 

At the radius $R_C=8.5$ kpc of the circular orbit considered in this study
we obtain $(\beta_C,\delta_C) = (1.37, 2.86)$ and the circular velocity $V_C=233.3$ km/s. 
The orbital time scale at $R_C=8.5$ kpc is $T_{\rm orb}\approx 224$ Myr.

\section{Cluster geometry}

According to King (1962), the projected density profile $\Sigma(r)$ of a star cluster
can be approximated by
\be
\Sigma(r) = k\left\{X^{-1/2}-C^{-1/2} \right\}^2 \label{eq:pd} \quad \mathrm{for} \quad r \le r_t
\ee
with normalisation constant $k$ and
\be
X(r, r_c) = 1+(r/r_c)^2 \quad \mathrm{and} \quad C(r_c, r_t) = 1+(r_t/r_c)^2 ,
\ee
where $r_c$ is the core radius and $r_t$ is the
(tidal) cutoff radius where the projected density of the model drops to zero. Integration yields the cumulative form of the King 1962 profile,
\bea
M_p(r) &=& 2\pi \int_0^{r} \Sigma(r') r' dr'  \quad \mathrm{for} \quad r \le r_t \nonumber \\
&=& \pi r_c^2 k \left\{\ln(X) - 4\frac{X^{1/2}-1}{C^{1/2}} + \frac{X-1}{C} \right\} \label{eq:pcm}
\eea
where $M_p(r)$ is the projected cumulative mass
of the model, i.e. the mass in projection on the sky within a circle of radius $r$.
For $r>r_t$ we force the integrated profile 
(Equation \ref{eq:pcm}) to the finite value
\be
M_p(r_t) = \pi r_c^2 k \left\{ \ln(C) - 3 + \frac{4}{C^{1/2}} - \frac{1}{C}. \right\}
\ee
We will not use the identification of $M_p(r_t)$ with the cluster mass, since in practice the equations are applied to star counts and the effective mass-of-light ratio enters the normalisation constant $k$.

\begin{figure*}
\includegraphics[width=1.0\textwidth]{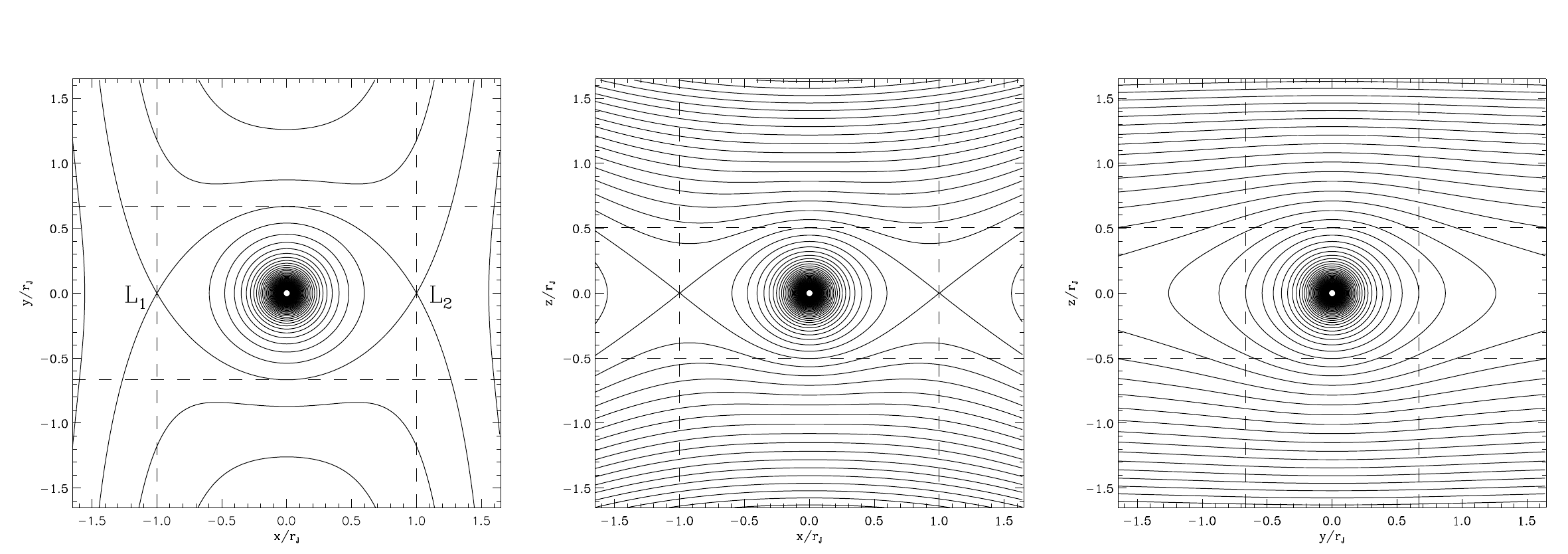} 
\caption{Cut through the equipotential surfaces of Equation (\ref{eq:phieff}) 
along the principal axis planes. The extents $x_{\rm max}$ (i.e. $r_J$), $y_{\rm max}$ (from Equation
\ref{eq:ymax}) and $z_{\rm max}$ (from Equation \ref{eq:zmax}) 
 of the last closed equipotential surface is marked with dashed lines. We assumed a Kepler potential for the cluster.} 
\label{fig:epkuzmin}
\end{figure*}

\begin{figure}
\includegraphics[angle=90,width=0.5\textwidth]{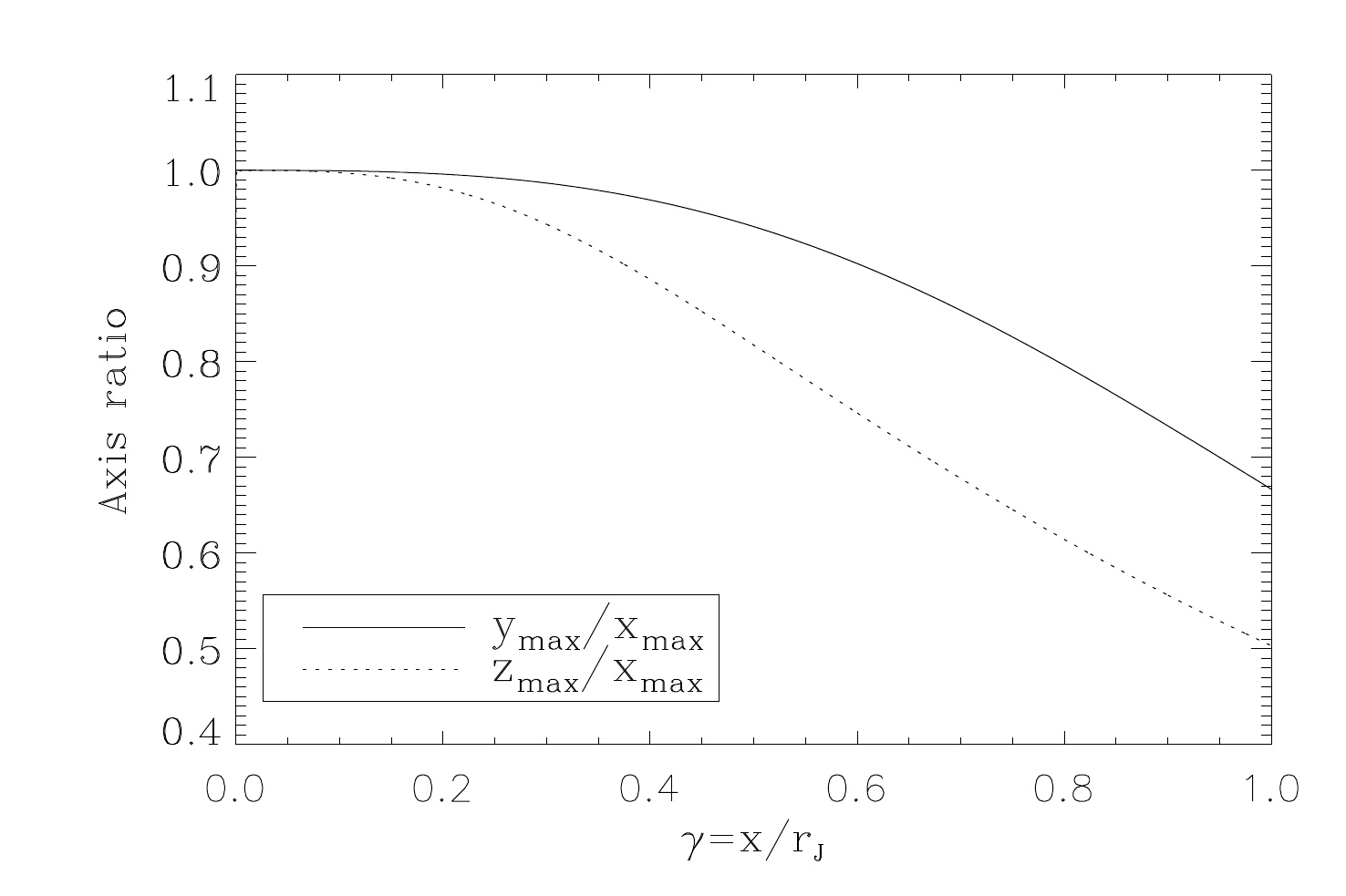} 
\caption{Principal axis ratios of the equipotential surfaces around the cluster as a function of
the parameter $\gamma=x/r_J$. At the centre of the cluster we have $\gamma=0$ while
$\gamma=1$ (dashed line) corresponds to the critical equipotential surface through $x=r_J$.
We assumed $(\beta,\delta)=(1.37,2.86)$ and a Kepler potential for the cluster.} 
\label{fig:ratio}
\end{figure}

Already King (1961) remarks that the tidal forces from the galaxy distort
only the outer regions of a star cluster. We quantify these deviations from
spherical symmetry due to the tidal field in this Section.
We employ a coordinate system (x,y,z) of ``principal axes of the star cluster''. Its
origin is the cluster centre. The x-axis points away from the galactic centre, the y-axis 
points in the direction of  the galactic rotation and the z-axis is directed towards 
the galactic north pole. 
Figure \ref{fig:epkuzmin} shows a ``principal axis plane cut'' through the equipotential 
surfaces of the effective potential around a star cluster on a circular orbit in the 
tidal field of the Milky Way. 
To second order, the effective potential is given by

\bea
\Phi_{\rm eff} &=& \Phi_{\rm eff,0} - \frac{GM_{\rm cl}}{\sqrt{x^2+y^2+z^2}} \nonumber \\
 && + \frac{1}{2} (\beta^2-4)\Omega^2 x^2 + \frac{1}{2} \delta^2 \Omega^2 z^2 \label{eq:phieff}
\eea

\noindent
For the cluster we assumed a Kepler 
potential, which is a very good approximation in the outer parts (Just et al. 2009). The unit in Figure \ref{fig:epkuzmin} is the Jacobi radius $r_J$. The Jacobi
radius is defined as the distance from the cluster centre to the Lagrange points
$L_1$ and $L_2$. It is given by

\be
r_J = \left[ \frac{GM_{\rm cl}}{(4-\beta^2)\Omega^2} \right]^{1/3} \label{eq:rjac}
\ee

\noindent
(see King 1962). The value of the effective potential on the critical equipotential surface which 
connects $L_1$ and $L_2$ can be easily calculated from (\ref{eq:phieff})
and (\ref{eq:rjac}). It is given by

\be
\Phi_{\rm eff,crit} = \Phi_{\rm eff,0} - \frac{3}{2} \frac{GM_{\rm cl}}{r_J}. \label{eq:phieffcrit}
\ee

\noindent
Assuming that $x_{\rm max}=r_J$, the radius in $y$-direction of the last closed (critical)
equipotential surface follows from (\ref{eq:phieff}) and (\ref{eq:phieffcrit}),

\be
y_{\rm max} = \frac{2}{3} r_J. \label{eq:ymax}
\ee

\noindent
For the radius in $z$-direction we have to solve a cubic equation

\be
-\frac{K}{z_{\rm max}} + \frac{1}{2} L^2 z_{\rm max}^2 = -\frac{3}{2} \frac{K}{M}
\ee

\noindent
with the constants

\be
K = GM_{\rm cl}, \ \ \ L = \delta\Omega, \ \ \ M = r_J.
\ee

\noindent
The only real solution is given by

\be
z_{\rm max} = \frac{\left\{ K \left[K+2L \left( LM^3 + \sqrt{KM^3+L^2M^6}\right)\right]\right\}^{1/3}-K^{2/3}}{L \left(LM^3 + \sqrt{KM^3+L^2M^6}\right)^{1/3} }
\ee

\noindent
For $(\beta,\delta)=(\beta_C, \delta_C)=(1.37,2.86)$ we obtain 
\be
z_{\rm max} \approx 0.503 \, r_J \label{eq:zmax}
\ee

\noindent
(e.g. Wielen 1974). More generally, the ratios of the principal axes
$y_{\rm max}:x_{\rm max}$ and $z_{\rm max}:x_{\rm max}$ of all
closed equipotential surfaces are shown
in Figure \ref{fig:ratio} as a function of the parameter $\gamma=x/r_J$.
At the centre of the cluster we have $\gamma=0$ while
$\gamma=1$ corresponds to the critical equipotential 
surface through $x=r_J$. 

\begin{figure*}
\includegraphics[width=1.0\textwidth]{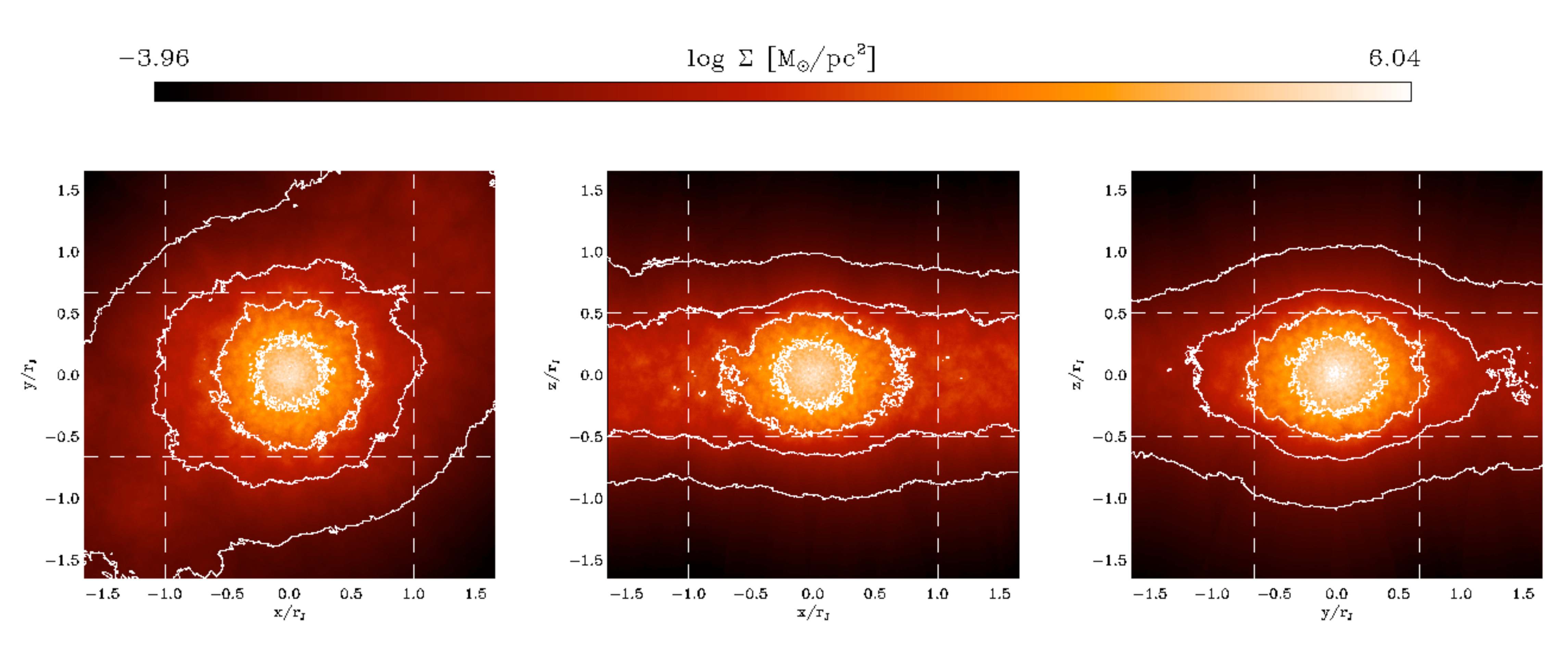} 
\caption{Surface density of projections onto the principal axis planes
of the cluster at $T_4=1.31$ Gyr. The dashed lines show the theoretical values of 
$x_{\rm max}/r_J$, $y_{\rm max}/r_J$ and $z_{\rm max}/r_J$. The contours
correspond to $\Delta\log\Sigma\approx 2$ dex.} 
\label{fig:paxissurf}
\end{figure*}

Figure \ref{fig:paxissurf} shows the logarithmically colour-coded
surface density $\Sigma$ of projections
of the simulated $N$-body model of the star cluster onto its principal axis 
planes at time $T_4=1.31$ Gyr. The surface density $\Sigma$ has been
calculated from the $N$-body snapshot file with the method of
Casertano \& Hut (1985) (their Equation II.6 with j=20).
The contours correspond to $\Delta\log\Sigma\approx 2$ dex.
The extent of the last closed (critical) equipotential surface
is marked with dashed lines. The
contours of constant surface density roughly follow the equipotential 
surfaces from Figure \ref{fig:epkuzmin}.

\section{Jacobi radius of the model}

We determined the Jacobi radius of our simulated $N$-body model iteratively from  Equation \ref{eq:rjac}.
The gravitational constant $G$ and the quantities $\beta$ and $\Omega$ are known  but $M_{\rm cl}$, the cluster mass within $r=r_J$ is unknown since $r_J$ is unknown. 
Our iteration is given by

\be
r_{n+1} =  \left[ \frac{GM_{\rm enc}(r_{n})}{(4-\beta^2)\Omega^2} \right]^{1/3} \label{eq:rjacit}
\ee

\noindent
where $M_{\rm enc}(r)$ is the enclosed mass of cluster stars within radius $r$
around the cluster centre.
Starting with $r_0=\infty$ the iteration theoretically converges towards $r_{\infty}=r_J$. 
In practice, a few iterations are sufficient to determine $r_J$ accurately. 
It is interesting to note that this simple method 
does not rely on numerical fits of the effective potential to the envelope of the spatial
distribution of Jacobi energies of the cluster stars. It can easily be applied to
an $N$-body snapshot file which contains masses and positions of the individual particles
at a certain time.

\section{Projection and fitting}

We calculate the projection at the sky in Galactic coordinates $(l,b)$ and
ignore here for simplicity the perspective effects.  
Strictly speaking the projection of the cluster model is consistent only in the
midplane $b=0^\circ$ corresponding to the circular orbit in the Galactic plane.  The rotation 
angle $\alpha$ is related to the galactic longitude $l$ by the law of sines,

\be
\sin\,\alpha = -\sin l\frac{R_0}{R_C}\quad R_C>R_0
\label{eq:alpha1}
\ee

\noindent
with $R_0=8$ kpc for the solar circle and $R_C=8.5$ kpc for the cluster orbit. At the Galactic coordinates (l,b) = (0,0) and (l,b)= (180$^\circ$,0) the ``system (x,y,z) of 
principal axes of the cluster'' with origin at the cluster centre is not rotated.
Note that at $l=90^\circ$ and
$l=270^\circ$ we have the maximum rotation angle $\alpha=-70^\circ .25$ and $\alpha=70^\circ .25$,
respectively.
\begin{figure}
\includegraphics[width=0.5\textwidth]{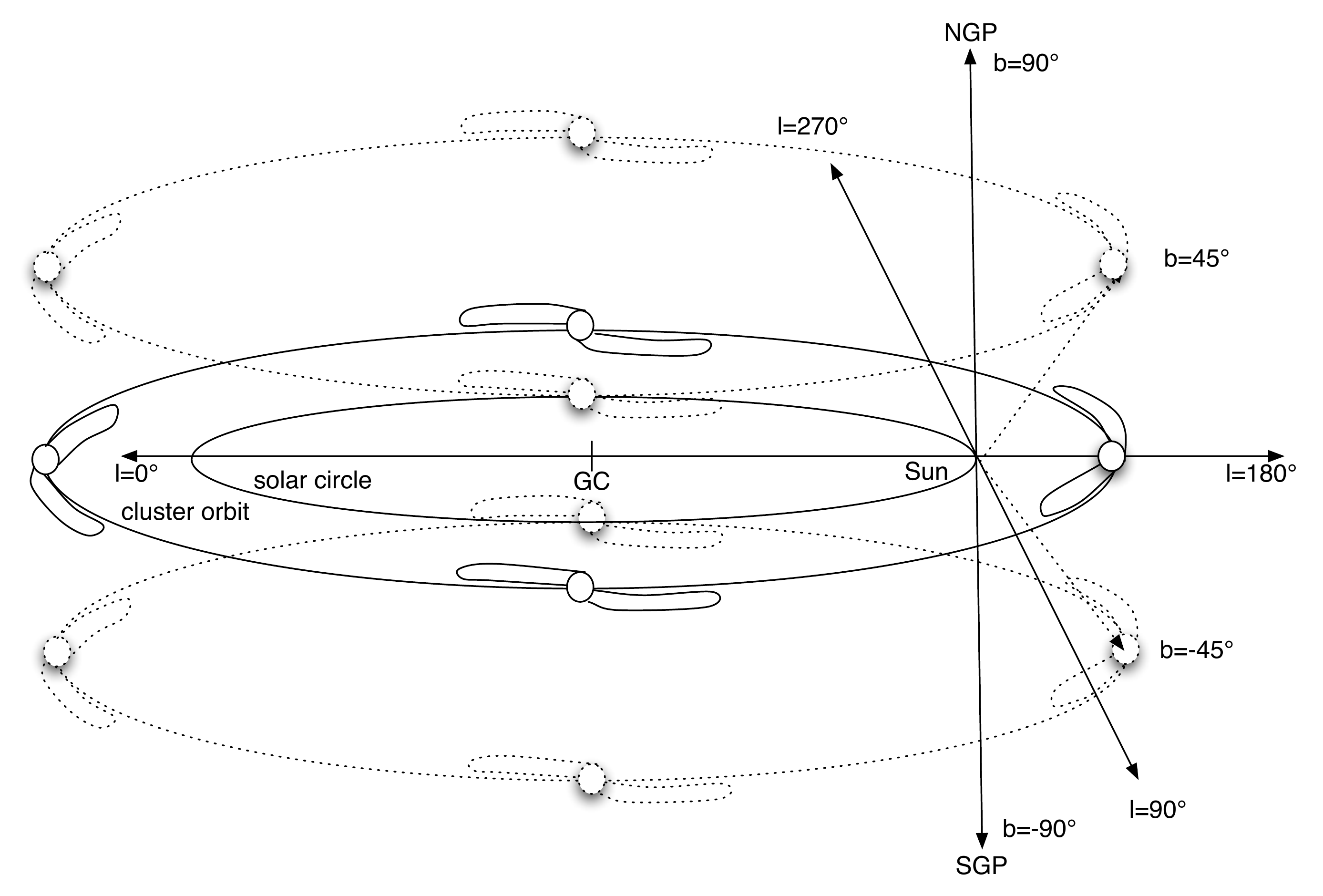} 
\caption{Sketch of the galactic coordinate system in which the projection is done.
The galactocentric radius of the sun is assumed to be $R_0=8.0$ kpc while the
cluster orbits at $R_C=8.5$ kpc on a circular orbit. 
The abbreviations denote the Galactic North (NGP)
and South (SGP) Pole and the Galactic Centre (GC).} 
\label{fig:galcoord}
\end{figure}

A sketch of the projection of the simulated $N$-body model of the star cluster
 onto the plane of the sky is illustrated in
Figure \ref{fig:galcoord}. Four orbital positions of the 
dissolving star cluster with its tidal tails are marked in the sketch. 

In order to demonstrate the effect of different projections also in galactic latitude, we add the cases of
a certain height of the cluster orbit above and below the plane of the solar circle (dotted lines in Figure \ref{fig:galcoord}).
The effect of the corresponding vertical oscillation of the cluster orbit on the intrinsic structure of the cluster and of a variation of $\delta$ in Equation \ref{eq:delta2} are neglected in this study. 


%

For all positions of the cluster on its orbit, we rotate the cluster around the $z$
axis and then around the $y'$ axis by the ordered pair of angles ($\alpha$, -b) in order 
to simulate the perspective of the cluster for an observer on earth.
After the projection, we determine the polar symmetric profile of the projected cumulative
mass $M_p(r)$ from our $N$-body data file by summations
over radius and polar angle and apply a fit with 
Equation \ref{eq:pcm}. For the fitting, we used the {\sc mpfit} package in
{\sc idl} (Markwardt 2009; Mor\'e 1978 for the Levenberg-Marquardt algorithm). 

\section{Results}

\begin{figure*}
\includegraphics[width=0.95\textwidth]{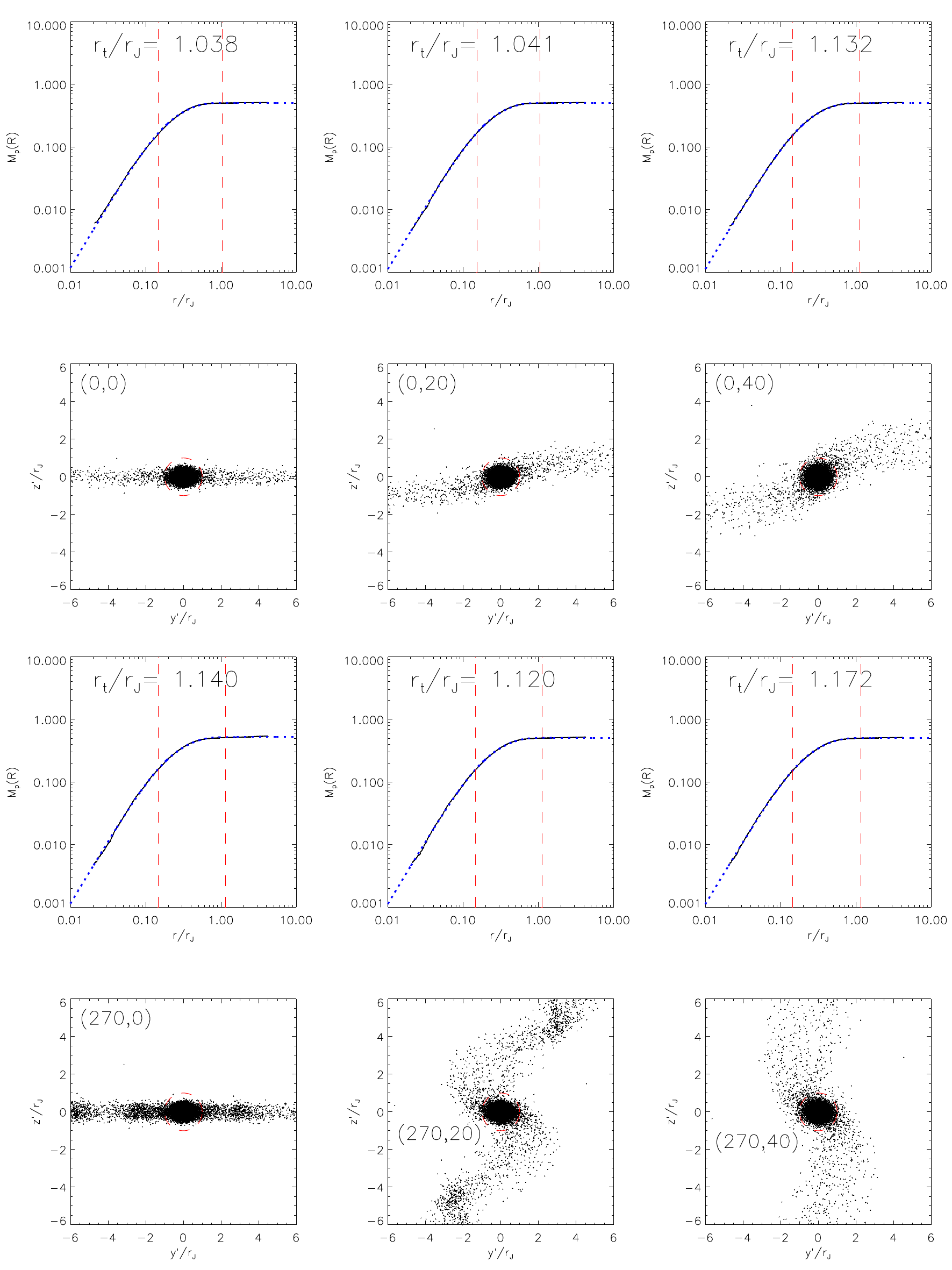} 
\caption{Examples of fits (upper panels) with the corresponding 
projections (lower panels) at time $T_4=1.31$ Gyr .
The resulting parameter ratio $r_t/r_J$ is given in the upper panels and the (l,b) coordinates
in the corresponding lower panels. In the upper panels, the solid (black) line 
represents the data and dotted (blue) line the fit. The dashed (red) lines mark $r_c$ 
(left dashed line) and $r_t$ (right dashed line) from the fit with Equation \ref{eq:pcm}.
In the lower panels, the dashed (red) line marks $r_J$.
} 
\label{fig:gridfit}
\end{figure*}

We first discuss projections according to Equation \ref{eq:alpha1} in detail. 
Figure \ref{fig:gridfit} shows examples of fits (upper panels) with the corresponding 
projections (lower panels). The resulting parameter ratio $r_t/r_J$ is given in the 
upper panels and the (l,b) coordinates in the corresponding lower panels. In the  upper panels, the solid (black) line represents the data and dotted (blue) line the fit.
The dashed (red) lines mark $r_c$ (left dashed line) and $r_t$ (right dashed line)
from the fit with Equation \ref{eq:pcm}. In the lower panels, the dashed (red) line 
marks $r_J$.

\begin{figure*}
\includegraphics[angle=90,width=1.0\textwidth]{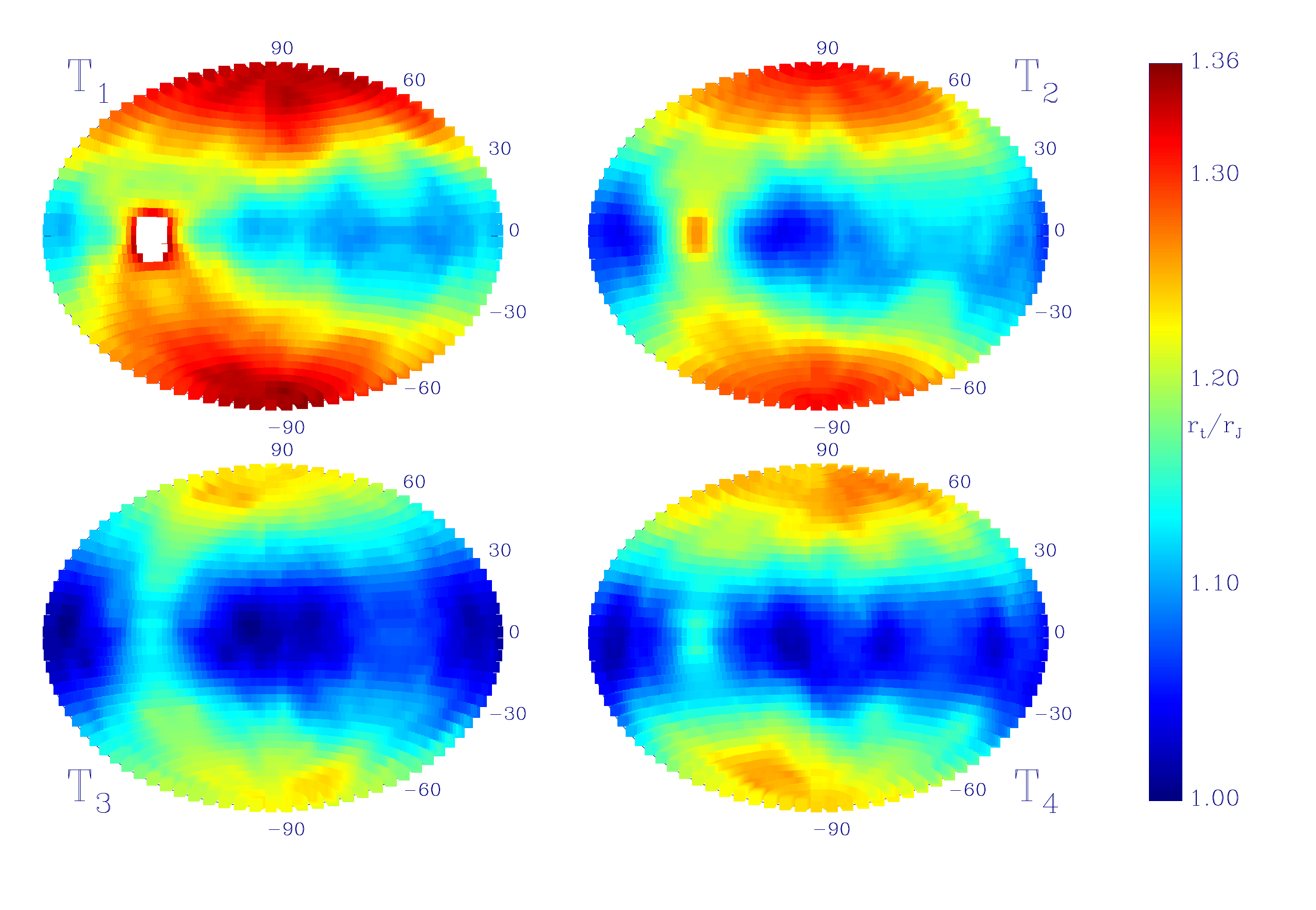} 
\caption{Parameter surfaces of $r_t/r_J$ as a function of Galactic coordinates for a 
fit of King (1962) models to projections on the sky of a simulated model at
different positions on its theoretical orbit. We used a squeezed Hammer-Aitoff projection. 
The color denotes the value of $r_t/r_J$ on a linear scale. 
The plots in the top row correspond
to $T_1=0.62$ Gyr (top left) and $T_2=0.84$ Gyr (top right). The plots in the bottom row
correspond to $T_3=1.06$ Gyr (bottom left) and $T_4=1.31$ Gyr (bottom right). 
} 
\label{fig:aisurf1}
\end{figure*}

\begin{figure}
\includegraphics[width=0.5\textwidth]{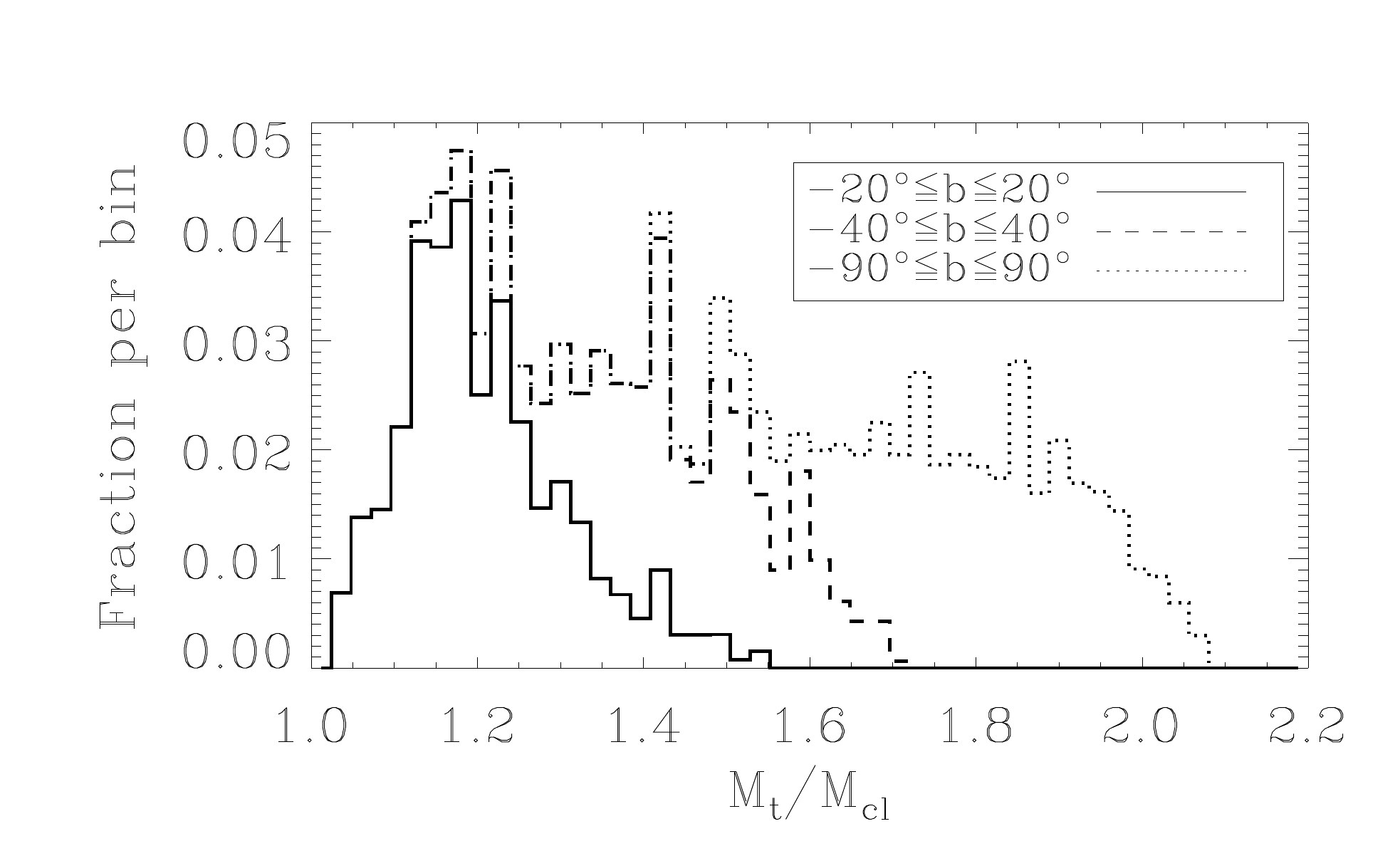} 
\caption{Histogram of the fraction on the sky per bin in $M_t/M_{cl}$
corresponding to the $T_4=1.31$ Gyr parameter surface from 
Figure \ref{fig:aisurf1}. $M_t$ and $M_{\rm cl}$ are the ``tidal masses'' 
calculated with Equation \ref{eq:rjac} from $r_t$ and $r_J$, respectively. 
} 
\label{fig:histrt}
\end{figure}

\begin{figure}
\includegraphics[angle=90,width=0.5\textwidth]{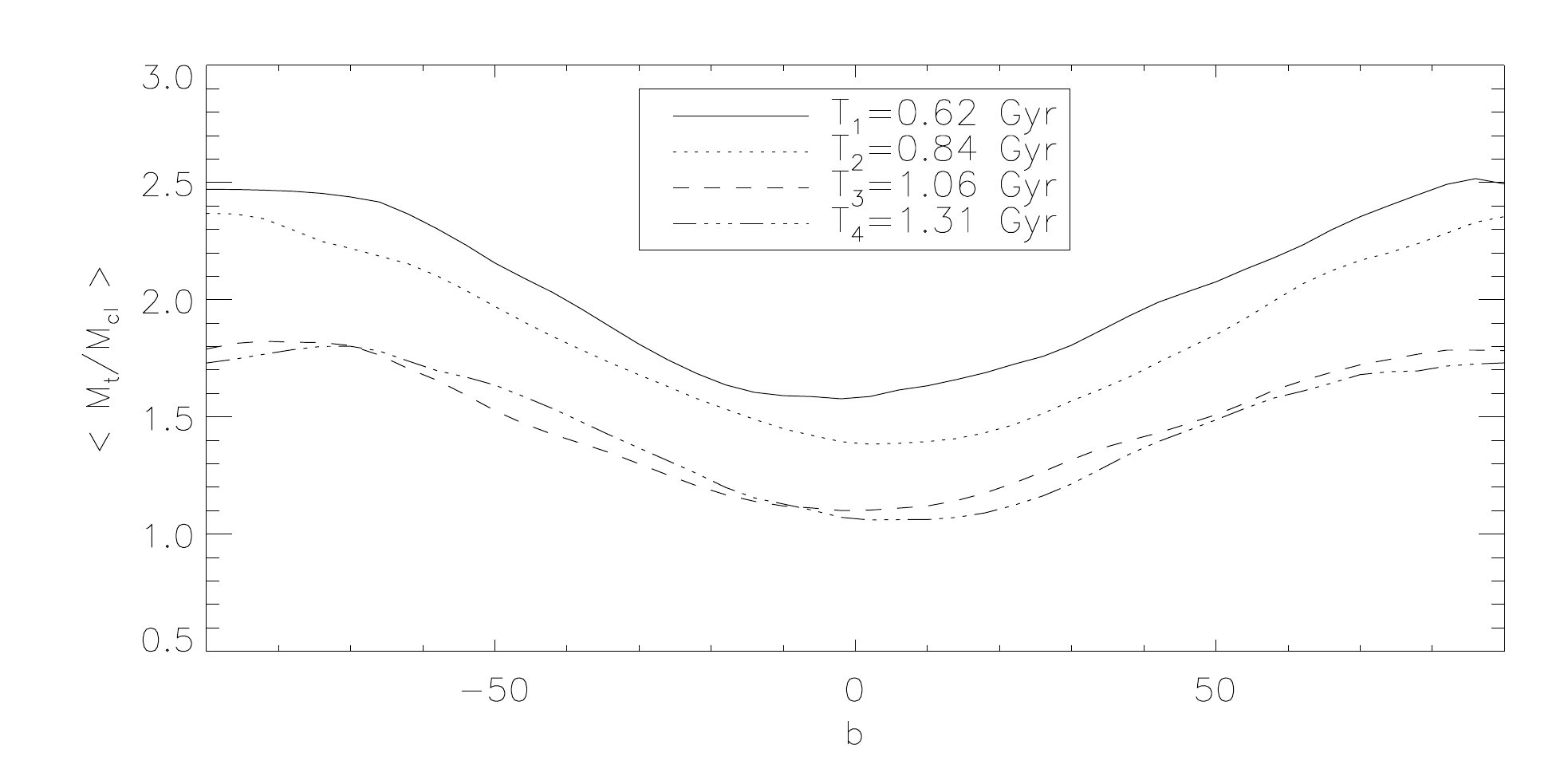} 
\caption{Mean mass $\langle M_t/M_{\rm cl} \rangle$ as a function of the galactic latitude $b$.
$M_t$ and $M_{\rm cl}$ are the ``tidal masses'' calculated with 
Equation \ref{eq:rjac} from $r_t$ and $r_J$, respectively. 
} 
\label{fig:meanmass}
\end{figure}

\begin{figure}
\includegraphics[angle=90,width=0.5\textwidth]{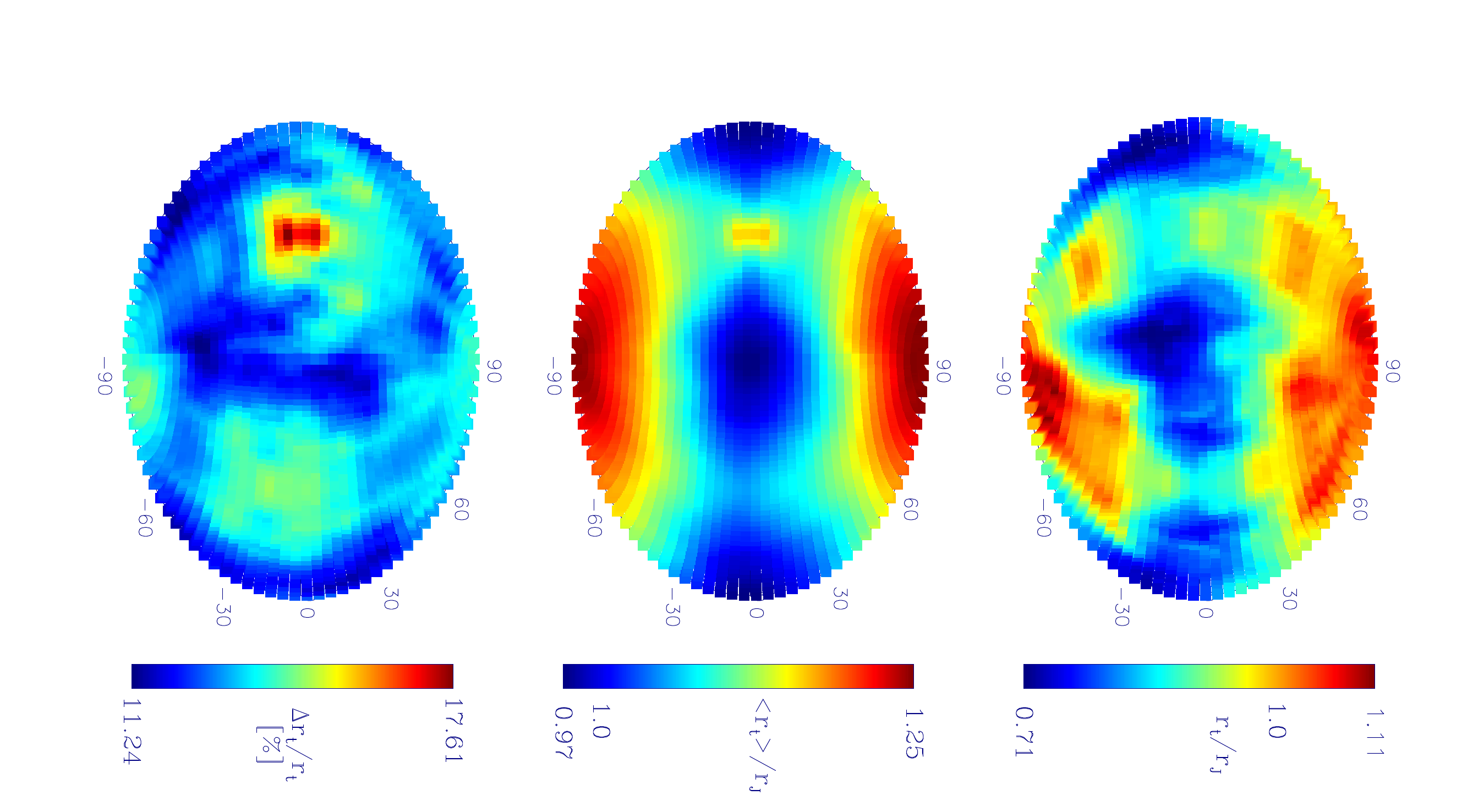} 
\caption{Parameter surfaces of $r_t/r_J$ and $\Delta r_t/r_t$ 
as a function of Galactic coordinates for a 
fit of King (1962) models to to projections on the sky of a simulated model at
different positions on its theoretical orbit. We used a squeezed Hammer-Aitoff projection. 
The time is $T_4=1.31$ Gyr. The color denotes the value of $r_t/r_J$  (and $\Delta r_t/r_t$, 
bottom plot) on a linear scale. 
The upper plots shows the parameter surface for the 400 brightest stars
in the simulated cluster (data courtesy of J. Beuria). The middle and bottom plots show the result 
of the bootstrap analysis (for explanations see the text). The middle plot 
shows the mean value of $r_t/r_J$ averaged over 100 small-number samples. 
The bottom plot shows the corresponding relative standard deviation $\Delta r_t/r_t$. 
} 
\label{fig:aisurf2}
\end{figure}

We derived the ratio $r_t/r_J$ for all projection directions in $(l,b)$ at four different evolution times 
of the cluster. The projections are done in steps of $4$ degrees from
$b=-90^\circ$ to $b=+90^\circ$ and $l=0$ to $l=360^\circ$. The full $N$-body snapshot file 
with $N=40404$ particles has been used for the projection and fitting procedure as described 
in Section 5. 
Figure \ref{fig:aisurf1} shows parameter surfaces 
of $r_t/r_J$ as a function of Galactic coordinates of the cluster centre at times
$T_1=0.62$ Gyr (top left), $T_2=0.84$ Gyr (top right), $T_3=1.06$ Gyr (bottom left)
and $T_4=1.31$ Gyr (bottom right). A squeezed Hammer-Aitoff projection of Galactic coordinates
has been used (see Appendix A). The total number of fits for each plot in Figure \ref{fig:aisurf1} is 
$N_{\rm fits}=4186$. Note that in the plot for $T_1$, a peak around $(l,b) \approx (270^\circ,0)$ 
(see also Figure \ref{fig:aiproj} in Appendix A) is not resolved (white colored area).
Here $r_t/r_J$ reaches a factor of 1.52.

It can be seen that the cluster masses are typically overestimated. 
The strongest bias to high masses at time $T_1$ (including the peak)
shows that the unbound stars stay for a long time close to the cluster
and contribute to the outer density profile in the fitting procedure. This effect depends 
on the initial conditions. At later times $T_3$, $T_4$ the cluster mass distribution becomes 
stable and the bias is independent of the age of the cluster.

Figure \ref{fig:histrt} shows a histogram of the fraction on the sky per bin
in $M_t/M_{cl}=(r_t/r_J)^3$ for the parameter surface in Figure \ref{fig:aisurf1} corresponding to
time $T_4$, where the ``tidal'' masses $M_t$ and $M_{\rm cl}$ are calculated with 
Equation \ref{eq:rjac} from $r_t$ and $r_J$, respectively. 
The solid line shows the distribution for projections
in the range between $-20^\circ < b < +20^\circ$. The most frequent
overestimation of the mass is $M_t/M_{cl}=1.18$. The dashed line shows the 
distribution for projections in the range between 
$-40^\circ < b < +40^\circ$. The most frequent ratio is the same as for 
the solid line. However, the ratios extend up to $M_t/M_{cl}=1.7$.
The dotted line shows the distribution for all projections.The most frequent ratio is the same as for 
the other two lines, but the ratios extend up to $M_t/M_{cl}\approx 2.1$.

Figure \ref{fig:meanmass} shows the mean mass $\langle M_t/M_{\rm cl} \rangle$ as a 
function of the galactic latitude $b$, where $M_t$ is the mass of cluster stars within radius $r_t$
which has been obtained by the fitting procedure. Shown are the curves corresponding to
times $T_1 - T_4$. All curves show a local minimum which is located roughly at $b=0^\circ$.
For time $T_1$ the mean overestimation of the mass reaches 2.5 in the direction 
of the Galactic poles.

Usually only a small fraction of cluster stars are identified as members leading to an increased statistical uncertainty in the fitting procedure. 
The uppermost plot in Figure \ref{fig:aisurf2} shows the parameter surface derived for the 400 most massive (i.e. the brightest) stars in the simulation at time $T_4$ (data courtesy of J. Beuria). 
The reason for the 
low values of $r_t/r_J$ is mass segregation. The stronger asymmetries in $\pm b$
compared to the plots in Figure \ref{fig:aisurf1} are due to a slightly asymmetric distribution of the 400 mass-segregated stars in position space.

In order to measure the statistical scatter in $r_t$ we applied a bootstrap analysis.
We divided the $N$-body snapshot file for time $T_4$ with $N=40404$ particles
into 100 small-number samples of $N_{\rm sample}=400$ particles each 
(the remaining particles were
dropped out of the analysis). For each of these small-number samples we applied the
procedure described above. The resulting total number of projections and fits 
was therefore $N_{\rm fits}=418600$. 
The middle and bottom plots in Figure \ref{fig:aisurf2} show the result of the bootstrap analysis.
The middle plot of Figure \ref{fig:aisurf2} shows the mean value of $r_t/r_J$ 
averaged over the 100 samples. 
The bottom plot shows the relative standard deviation $\Delta r_t/r_t$ which resulted 
from the averaging over the 100 samples. The uncertainty on a single determination 
of $r_t$ lies in the range between $10$ and $20$ percent. The highest
uncertainty is expected for $r_t$-determinations in the vicinity of the peak 
around $(l,b) \approx (270^\circ,0)$.

A comparison with the uppermost plot in Figure \ref{fig:aisurf2} 
shows that the derived limiting radii for the 
sample of the most massive stars are systematically lower due to mass segregation. 
The differences are typically on a $1-2\sigma$ level.



\section{Conclusions}

The result of the analysis in this paper is the confirmation that the star 
cluster masses are typically overestimated if the method by Piskunov 
et al. (2007) is applied on a complete sample.
Moreover, we quantified the methodological error in our analysis.

Figures \ref{fig:aisurf1} and \ref{fig:aisurf2} show that at certain Galactic coordinates the
King (1962) profile fits are particularly biased. A high bias is predicted
for $(l,b) \approx (270^\circ,0)$ (see Figures \ref{fig:aisurf1}, \ref{fig:aisurf2} and \ref{fig:aiproj}). 
The corresponding rotation angle of the cluster is $\alpha=70^\circ$, where the projection is parallel to the inner end of the tidal arms (see lower left plots in Figure \ref{fig:gridfit}). 
For $(l,b) = (90^\circ,0)$ there is no
corresponding peak in the parameter surface (best visible for $T_1$)
due to the asymmetry between the leading and trailing tidal tails (see the
sketch in Figure \ref{fig:galcoord}).
Also, for Galactic latitudes beyond $b\approx \pm 40^\circ$ (cf. Figure \ref{fig:aiproj})
the bias becomes large but only few OCs are located in that regime.

For the parameter surfaces in Figure \ref{fig:aisurf1} the
masses are biased within the ranges 
[$1.3 \, M_{\rm cl}$, $3.5 \, M_{\rm cl}$] (at $T_1$), 
[$1.1 \, M_{\rm cl}$, $2.3 \, M_{\rm cl}$] (at $T_2$), 
[$1.0 \, M_{\rm cl}$, $2.0 \, M_{\rm cl}$] (at $T_3$) and 
[$1.0 \, M_{\rm cl}$, $2.1\, M_{\rm cl}$] (at $T_4$) 
depending on the evolutionary state of the star cluster in the tidal field of the Galaxy.
The bias depends strongly on the projection angles, which transform differently to Galactic coordinates for different orbital radii of the OC.

Furthermore, a bootstrap analysis showed that 
the relative error on a single determination of the limiting radius $r_t$ lies in the range 
between $10$ and $20$ percent (at time $T_4$) corresponding to an uncertainty in the mass of $\approx 50$ percent for samples
of $N_{\rm sample}=400$ particles (which are typical for rich OCs). 

Mass segregation of the brightest stars in a cluster 
can alter the $r_t/r_J$ factor significantly which is important
for the data analysis of observations. The mass segregation results in a
concentrated core which leads to an underestimation of the tidal radius.
For a younger cluster age one would expect lesser mass segregation.

For a quantitative correction of the bias in the cluster mass determination by identifying $r_t$ with $r_J$ an extensive parameter study of cluster parameters is necessary. The influence of mass-segregation on selection effects concerning the brightness limit of the observations should be included, because stellar evolution is taken into account. Especially for young clusters the bias factor can be sensitive to the initial conditions. The final goal is to find an agreement of OC mass determinations by the different methods. This would allow an interesting insight in the OC properties like the IMF, mass-to-light ratio and mass segregation. 

\section{Acknowledgements}


The authors thank J. Beuria for his help with the Hammer-Aitoff projections and
the provision of the data for the uppermost plot in Figure \ref{fig:aisurf2}.

PB and MIP acknowledge the special support by the
Ukrainian National Academy of Sciences under the
Main Astronomical Observatory GRAPE/GRID computing
cluster project.

PB acknowledges the support from the Volkswagen
Foundation GRACE Project No. I80 041-043.

 MIP acknowledges support by the University of Vienna through
the frame of the Initiative Kolleg (IK) `The Cosmic Matter Circuit'
I033-N and computing time on the Grape Cluster of the University
of Vienna.

\appendix

\section{Squeezed Hammer-Aitoff projection}

\begin{figure}
\includegraphics[angle=90,width=0.5\textwidth]{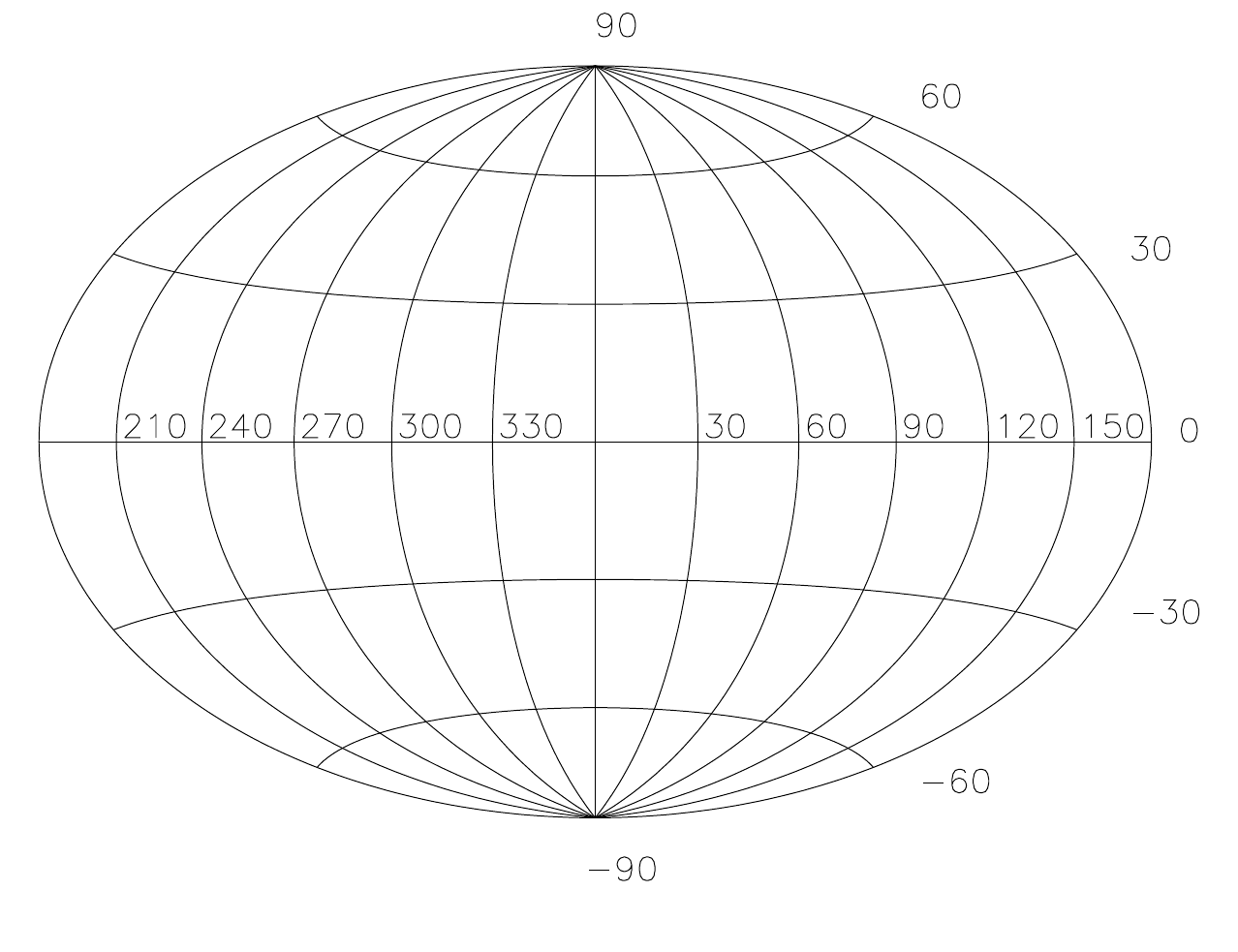} 
\caption{Squeezed Hammer-Aitoff projection. The Galactic latitude runs from
$b=-90^\circ$ to $b=+90^\circ$ and the longitude from $l=0$ to $l=360^\circ$.
} 
\label{fig:aiproj}
\end{figure}

The squeezed Hammer-Aitoff projection is given by

\bea
x &=& 2f\frac{\cos(b)\sin(l/2)}{\sqrt{1+\cos(b)\cos(l/2)}}, \label{eq:a1} \\
y &=& \frac{2}{f}\frac{\sin(b)}{\sqrt{1+\cos(b)\cos(l/2)}}. \label{eq:a2}
\eea

\noindent
It is the standard Hammer-Aitoff equal-area projection where we introduced a 
free squeezing factor $f$. The squeezing leaves the area element 
$dA=dxdy=\sqrt{\mathrm{det} g} \, db dl  = \cos(b)dbdl$ invariant,
where $g$ is the first fundamental form calculated from equations (\ref{eq:a1})
and (\ref{eq:a2}). The ratio of diameters of the elliptic projection area
is given by $d_x/d_y=f^2$.
In the standard Hammer-Aitoff projection we have $f=\sqrt{2}$. For a projection
onto a circular area one can set $f=1$. 
The inverse projection is given by

\bea
z &=& \sqrt{2-\left(\frac{x}{2f}\right)^2-\left(\frac{fy}{2}\right)^2} \label{eq:auxil} \\
b &=& \arcsin\left(\frac{f}{2}zy\right), \\
l &=& 2\arctan\left(\frac{1}{2f}\frac{zx}{z^2-1}\right),
\eea

\noindent
where we have introduced the auxiliary variable $z$.

We used $f=1.2$ for Figures \ref{fig:aisurf1} and \ref{fig:aisurf2}.
The resulting coordinate system  (which is hidden in Figures \ref{fig:aisurf1} and 
\ref{fig:aisurf2}) can be seen in Figure \ref{fig:aiproj}. We have modified {\sc idl}
routines by W. B. Landsman to incorporate the free squeezing factor.


\begin{thebibliography}{99}

\bibitem[Aarseth 1999]{} Aarseth S. J., Publ. Astron. Soc. Pacific 111, 1333 (1999)
\bibitem[Aarseth 2003]{} Aarseth S. J., {\it Gravitational $N$-body simulations --
Tools and Algorithms}, Cambridge Univ. Press (2003)
\bibitem[Casertano 1985]{} Casertano S., Hut P., Ap. J., 298, 80
\bibitem[Dauphole 1995]{} Dauphole B., Colin J., 1995, A\&A 300, 117
\bibitem[Harfst 2007]{} Harfst S., Gualandris A., Merritt D., Spurzem R., Portegies Zwart S., 
Berczik P., 2007, New Astron., 12, 357 
\bibitem[Just 2009]{} Just A., Berczik P., Petrov M. I., Ernst A., 2009, MNRAS, 392, 969
\bibitem[Kharchenko 2009], Kharchenko N. V., Berczik P., Petrov M. I., Piskunov A. E., R\"oser S., Schilbach E., Scholz R.-D., 2009, A\&A 495, 807
\bibitem[King 1961]{} King I., 1961, AJ 66, 68
\bibitem[King 1962]{} King I., 1962, AJ 67, 471
\bibitem[Makino 1992]{} Makino J., Aarseth S. J., 1992, PASJ, 44, 141 
\bibitem[Markwardt 2009]{} Markwardt C. B., 2009, in proc. Astronomical Data Analysis Software and Systems XVIII, Quebec, Canada, ASP Conference Series, Vol. 411, eds. D. Bohlender, P. Dowler \& D. Durand, Astronomical Society of the Pacific, San Francisco, p. 251-254 
\bibitem[Miller 1978]{} Miller G.E \& Scalo J. M., 1978, PASP, 90, 506
\bibitem[Miyamoto 1975]{} Miyamoto M. \& Nagai R., 1975, PASJ, 27, 533
\bibitem[More 1978]{} Mor\'e J., 1978, in Numerical Analysis, vol. 630, ed. G. A. Watson, 
Springer Verlag, Berlin, p. 105
\bibitem[Oort 1965]{} Oort J. H., in Blaauw A., Schmidt M., eds, Galactic Structure,
Univ. Chicago Press, Chicago, IL, p. 455
\bibitem[Piskunov 2007]{} Piskunov A. E., Schilbach E., Kharchenko N. V., R\"oser S.,
Scholz R.-D., 2007, A\&A 468, 151
\bibitem[Piskunov 2008a]{} Piskunov A. E., Schilbach E., Kharchenko N. V., R\"oser S., Scholz R.-D., 2008a, A\&A 477, 165
\bibitem[Piskunov 2008b]{} Piskunov A. E., Kharchenko N. V., Schilbach E., R\"oser S., Scholz R.-D., Zinnecker, H., 2008b, A\&A 487, 557
\bibitem[Roeser 2010]{} Roeser S., Kharchenko N. V., Piskunov A. E., Schilbach E.,
Scholz R.-D., Zinnecker H., 2010, Astron. Nachr. 331, 519 
\bibitem[Spurzem 1999]{} Spurzem R., J. Comp. Applied Maths. 109, 407 (1999) 
\bibitem[Wielen 1971]{} Wielen R., 1971, A\&A 13, 309
 \bibitem[Wielen 1974]{} Wielen R., 1974, in Mavridis L. N, ed., Proceedings of the 
1st European Astronomical Meeting, Vol. 2, Stars and the Milky Way System,
Springer, Berlin, p. 326


\end{thebibliography}
\end{document}